# A Wearable IoT Aldehyde Sensor for Pediatric Asthma Research and Management

Baichen Li, Quan Dong, R. Scott Downen, Nam Tran, J. Hunter Jackson, Dinesh Pillai, Mona Zaghloul, Zhenyu Li[*]

**Abstract**

Mechanistic studies of pediatric asthma require objective measures of environmental exposure metrics correlated with physiological responses. Here we report a cloud-based wearable IoT sensor system which can measure an asthma patient's exposure to aldehydes, a known class of airway irritants, in real-life settings. The wrist-watch shaped sensor can measure formaldehyde levels in air from 30ppb to 10ppm using fuel cell technology, and continuously operate over 7 days without recharging. The sensor wirelessly uploads data to an Android smartphone via Bluetooth Low Energy (BLE). The smartphone also functions as a gateway to a cloud-based informatics system which handles sensor data storage, management and analytics. Potential applications of this IoT sensor system include epidemiological studies of asthma development and exacerbations, personalized asthma management and environmental monitoring.

## 1. Introduction

Asthma is one of the most common chronic diseases in children, affecting approximately 6.1 million children under 18 in the US alone (8.3% of the pediatric population) [1]. It is estimated to be the leading cause of missed school days and the third-ranking cause of hospitalization for children [2]. The annual cost of treating pediatric asthma is $3.2 billion dollars [3]. Triggers of pediatric asthma exacerbations such as air pollution, tobacco smoke, allergies, and airway infections are well known. However, how the interaction between environmental exposure and the patient's biological response determines susceptibility and timing of such events is less clear. One significant barrier to causal understanding is the lack of objective measures of exposure metrics correlated with patient physiological responses and activities.

Air pollution is known to be associated with asthma exacerbations [4–6]. For example, formaldehyde, a common air pollutant known to irritate airways, is found in tobacco smoke, furniture, paintings, synthetic carpets, and other everyday sources [7–13]. Previous studies have suggested a significant positive association between formaldehyde exposure and asthma exacerbations [14–22]. However, currently no sensing system is available to researchers which can continuously monitor an individual patient's exposure to formaldehyde and correlate it with the patient's symptoms and physiological responses. Formaldehyde sensors based on metal oxide technology and colorimetric assays have been developed before [23–25], but not in wearable formats suitable for real-life monitoring. In addition, to make large-scale epidemiological studies feasible, it's likely a cloud-based IoT-like system is required to handle large-scale sensor deployment, management, data storage, and analytics [26–34].

To address this unmet need, we have developed a wearable IoT aldehyde sensor integrated with a cloud-based informatics system. The wearable IoT aldehyde sensor is based on electrochemical fuel cell technology and can measure formaldehyde in air at levels as low as 30ppb, below the OSHA lower limit of 0.75ppm (8 hour time-weighted average) [35,36]. The final device resembles a wrist watch with dimensions of 36mm × 42mm × 11mm and a weight of 35g. In use, the device can continuously operate over 7 days without recharging. The built-in Bluetooth Low Energy (BLE) capability allows it to wirelessly upload data to an Android smartphone and display data in real time. The smartphone also functions as a gateway to an Amazon cloud-based informatics system which handles sensor data storage, management and analytics. Asthma researchers and physicians can visualize and analyze the sensor data through a web browser.

**Material and Methods**

*1.1  System Overview*

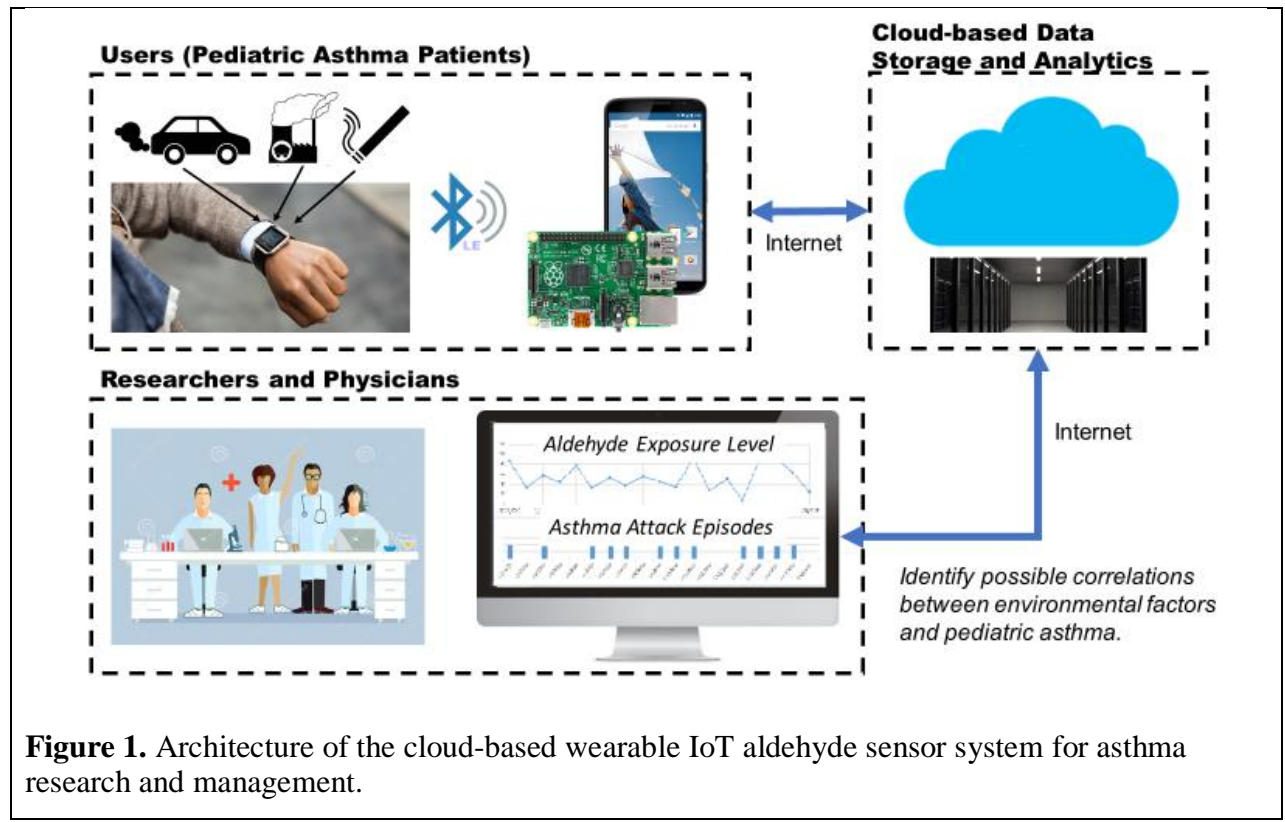

**Figure 1.** Architecture of the cloud-based wearable IoT aldehyde sensor system for asthma research and management.

The architecture of the system is shown in **Figure 1**, which consists of three major components: (1) a wearable sensor device on the wrist, (2) a BLE-capable gateway device with internet connectivity (such as a smartphone or a computer), and (3) a cloud-based data analytics system.

The wearable device utilizes an electrochemical fuel cell for aldehyde sensing and a digital CMOS sensor for temperature and humidity monitoring. Low-power, miniaturized electronic components are employed to build a self-contained data acquisition and transmission module, including an analog front-end (AFE), a BLE-capable wireless microcontroller, flash memory, power management circuitry and a lithium battery.

The wearable sensing device was designed to resemble a wrist watch. To minimize the need for user intervention, no buttons or indicators are included in the design. When powered on, the device will continuously measure ambient environmental conditions at a sampling interval of 1 dataset per second (aldehyde level, temperature, relative humidity, etc.). Measurement data is then averaged every minute and stored to flash memory, which can then be uploaded to the cloud via a BLE-capable gateway device using a custom software communication protocol.

An Android-based application was developed on a Google Nexus 5X smartphone serving as a gateway. Once a BLE connection is established between the application and the wearable sensing device, measurement data stored on the device is retrieved. Real-time measurement data can also be continuously transferred and displayed on the smartphone.

If internet connectivity is available, data stored on the central BLE device is uploaded to a cloud-based informatics system for centralized data storage, visualization and analysis. This cloud-based data analytics system is implemented on Amazon Web Services (AWS) with a simple dashboard configured to demonstrate the basic data analytics capabilities. Users can access this cloud-based analytics system via a secure web interface. Time-stamped measurement data can also be downloaded to a local computer for quantitative analysis using data analysis software (e.g. MATLAB, R, SPSS).

*1.2   Wearable Sensor Specifications*

The U.S. Environmental Protection Agency (EPA) estimates that the average in-home formaldehyde exposure levels range from 0.10 to 3.8 parts per million (ppm), while outdoor levels in U.S. urban area are in the range of 11 to 20 parts per billion (ppb) [37]. The Occupational Safety and Health Administration (OSHA) requires work place exposures to formaldehyde to be less than 0.75ppm as an eight-hour time-weighted average (TWA) [35]. Studies suggest that the time-weighted average of indoor formaldehyde exposure levels should be limited to 0.1 parts-per-million (ppm) [38]. Based on these guidelines, the target detection range for aldehyde was set to be 30 ppb to 10 ppm.

Certain weather conditions are also significant asthma triggers, including high relative humidity (RH) and freezing temperatures [39,40]. Moreover, it is important to calibrate the aldehyde fuel cell sensor output against temperature and humidity levels, as the sensor response can be affected by these factors. Therefore, a calibrated digital temperature and RH sensor was also included.

A self-contained system with a sturdy package is necessary for a wearable sensor to be used in real-world environments, with modern smart watches and wristbands serving as a good reference model. For end-user convenience, it is desirable to make such devices rechargeable with a battery life longer than 24 hours. Also, the wearable sensor should be as small and light as possible, especially for pediatric users. For these reasons, all internal components were packaged into the size of an Apple watch (36 x 42 x 11mm). A USB 2.0 micro-B charging port was also added which is compatible with standard USB chargers (5.5V).

Bluetooth Low-Energy (BLE) is a low-power wireless data transmission technology widely adopted by smartphones and IoT devices. It is currently estimated that billions of devices are BLE-ready across the globe [41]. With the help of a custom smartphone application, end users will be able to view sensor data in real-time and contribute to research by sharing their data. Moreover, readily available commercial microcontrollers with BLE functions have been developed for emerging IoT devices. A low-power BLE-enabled, small-footprint microcontroller (Texas Instruments, CC2650F128RSM) was selected for our wearable sensor platform.

*1.3   Fuel Cell Aldehyde Sensor*

A fuel cell aldehyde sensor is a two-electrode electrochemical sensor, which converts chemical energy into electrical current. The energy conversion is accomplished by a sandwich wafer structure (**Figure 2**): two outer layers (anode and cathode) of a fuel cell sensor are coated with Platinum (Pt) catalyst materials and an intermediate electrolyte layer is constructed for protons to pass through [42]. Aldehyde is oxidized at the anode while oxygen is reduced at the cathode. The generated protons travel through the intermediate layer (a dilute aqueous acidic electrolyte soaked into a porous plastic separator) to reach the cathode side and react with oxygen. If a load is connected between the anode and cathode, the generated electrons will flow through the external circuit to neutralize the system and complete the reaction. Consequently, an electrical current is generated.

The electrode reactions in a low temperature formaldehyde/Pt fuel-cell can be expressed as follows [43]:

Anode: $HCHO + H_2O - 2e^- \rightarrow HCOOH + 2H^+$

Cathode: $2H^+ + (½)O_2 + 2e^- \rightarrow H_2O$

According to the manufacturer, the fuel cell sensor used in this work (Dart Sensors 2-FE5) yields two electrons for every formaldehyde molecule consumed, i.e. the anode reaction is likely not complete and generates formic acid instead of $CO_2$. The reaction mechanism likely works on the methylene glycol molecule which is formed when formaldehyde dissolves in water. Each methylene glycol molecule contains two (OH) groups each releasing a hydrogen atom to the catalyst to react.

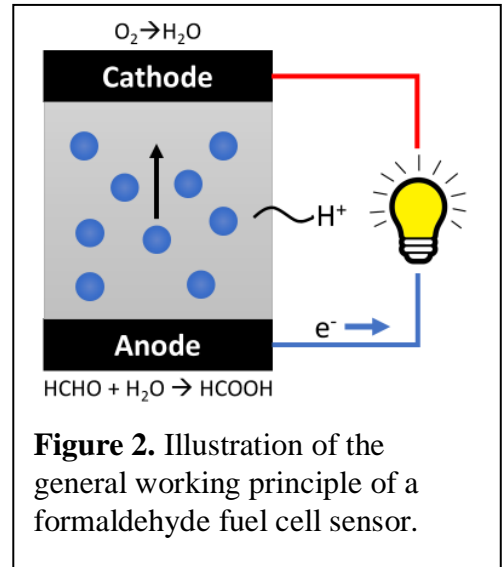

**Figure 2.** Illustration of the general working principle of a formaldehyde fuel cell sensor.

This wearable sensor operates in the diffusion regime, thus no sampling actuator (e.g. a pump) is required. Aldehyde gas present in the ambient environment diffuses into the fuel cell through an opening (1mm in diameter) on the sensor package. A porous hydrophobic membrane filter is placed above the opening to prevent liquid from getting into the sensor.

### 1.4 Temperature and Humidity Sensor

A fully calibrated digital temperature and relative humidity sensor (HTU21D) from TE Measurement Specialties is employed in the wearable sensor to continuously monitor the temperature and relative humidity levels in the ambient environment. Temperature and relative humidity are measured by internal transducers with accuracies of ±0.3ºC and ±2%, respectively. The output signals are digitized by a built-in ADC and stored in the internal registers of the chip.

### 1.5 Electronic System Design

**Figure 2** shows the complete block diagram of the electronic system design. An ultra-low power BLE-compatible microcontroller (Texas Instruments, CC2650F128RSM) with a footprint of 4 x 4 mm was used to design and prototype the wearable sensor. The CC2650 microcontroller contains a 32-bit ARM® Cortex®-M3 processor operating at 48 MHz as the main processor, many built-in peripherals, and an ultra-low power sensor controller. This sensor controller enables considerable power saving by interfacing external sensors and collecting data autonomously while the rest of the system is in sleep mode.

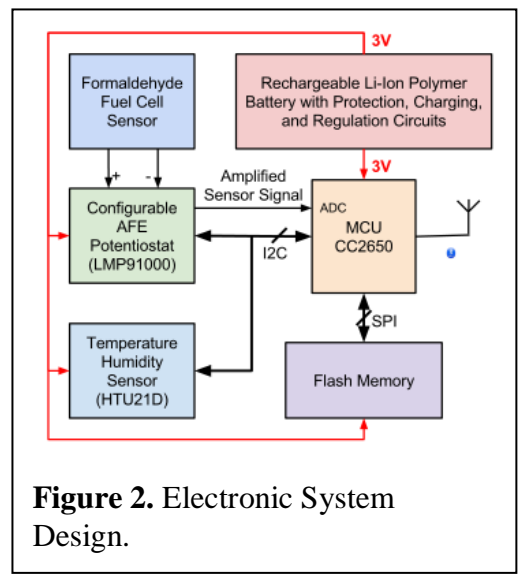

**Figure 2.** Electronic System Design.

A low-power configurable AFE potentiostat chip (LMP91000) was employed to interface between the formaldehyde sensor and the built-in ADC in the microcontroller. This chip is a potentiostat designed to support various 3-lead and 2-lead electrochemical sensors and can be further configured by a microcontroller. The electrical current from the formaldehyde fuel cell sensor is amplified by the internal TIA with programmable gains. The amplified output voltage is then quantized by the built-in ADC of the microcontroller. This potentiostat chip can be easily configured to adapt to other electrochemical sensors in the future.

A low-power 1 megabyte flash memory device (MX25R8035F) was added to the system for data storage. Like other flash memories of this style, the memory spaces are divided into pages of 256 bytes. The first 16 bytes in a page are used to store the metadata (e.g. page ID, microcontroller time, UTC time, etc.), and the following 240 bytes to store the measurement data. At a sampling interval of 1 dataset/min, this flash memory can store over 100 days of data, which is sufficient for current applications. In the case the memory is full, the program wraps the writing pointer to 0 and keeps writing from the beginning of the memory.

All electronic components were powered through a 3-volt voltage regulator (TPS78230) powered by a rechargeable lithium-ion polymer battery (DTP4015245) with built-in protection circuitry. Moreover, a lithium-ion polymer battery charging controller (MCP73831) was connected between the USB charging port and the lithium battery for safe charging.

## 1.6 Firmware Design

Besides the transducing mechanisms, two essential functions are required for a practical wearable sensing device: (1) sensor data acquisition and storage and (2) real-time and historical measurement data retrieval. The firmware was developed on TI BLE-Stack

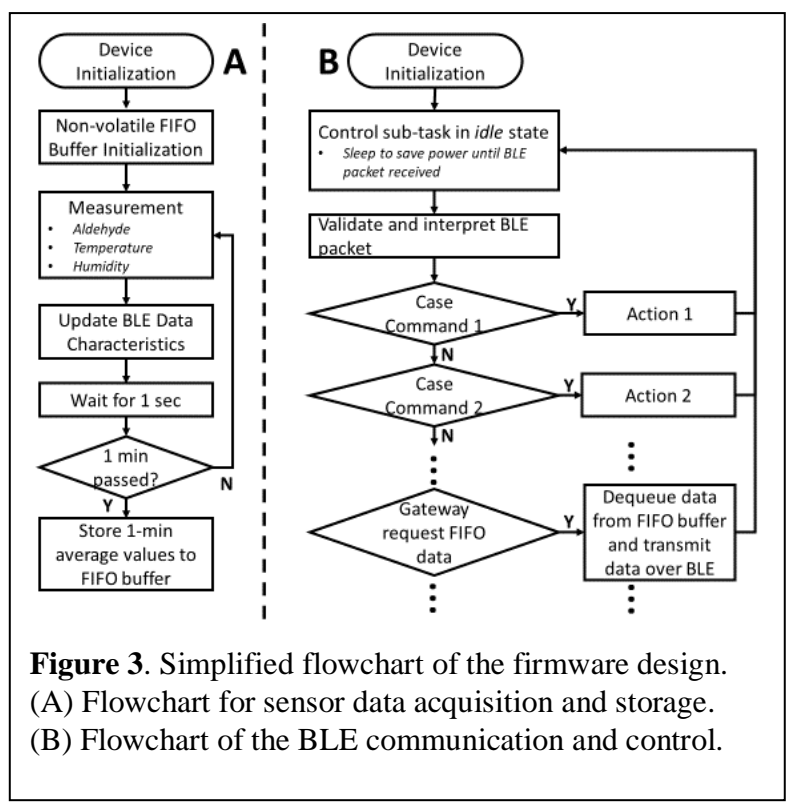

**Figure 3**. Simplified flowchart of the firmware design. (A) Flowchart for sensor data acquisition and storage. (B) Flowchart of the BLE communication and control.

v2.2.x SDK with TI's real-time operating system (RTOS). In the data acquisition task (**Figure 3.A**), measurement data was taken every second and stored in the on-chip random-access memory (RAM). Both digital (*temperature and relative humidity*) and analog (*aldehyde fuel cell*) sensors were used in this wearable device. Every second, the temperature and humidity sensor measurements were acquired, while the output current of the aldehyde fuel cell was amplified by a transimpedance amplifier and digitized with the built-in 12-bit ADC on the microcontroller.

The external flash memory was managed as a non-volatile FIFO buffer, where measurement data can be stored. To reduce memory usage, 1-minute consecutive measurement data was averaged before it was saved in the FIFO buffer. Additionally, local and UTC timestamps can be transmitted to the wearable sensor and stored in the FIFO buffer from a gateway device for synchronization. Historical data stored in the FIFO buffer can be retrieved by a gateway device using custom BLE commands (**Figure 3.B**).

*1.7   Bluetooth-Low Energy Communication Protocols*

BLE devices communicate using the Generic Attributes (GATT) Profile protocol, which is based on the Attribute protocol (ATT). Sensor information and measurement data are divided and encapsulated in logical entities called *characteristics*, and multiple characteristics are grouped into a *service* to serve a specific purpose.

A simple service for communication between wearable sensors (BLE peripheral) and a gateway device (central) was designed and implemented. It consisted of five characteristics: one for transmitting general-purpose data (TX), one for receiving general-purpose data (RX), and the other three (DATA) contain sensor measurement data. TX and RX characteristics were configured to be 20-bytes in length, the maximum value allowed in BLE. A custom packet-based communication protocol was designed for sensor configuration and historical data retrieval. Data characteristics were defined to store measurement data (temperature, relative humidity, and aldehyde level) in single-precision floating-point format (4-bytes) updated every second. In future developments, the number of data characteristics can be increased and the sampling frequency can be reconfigured based on the requirements to support other sensing modalities.

*1.8   Aldehyde Sensor Calibration*

To evaluate the responsivity and baseline at different temperatures, a calibration experiment setup (**Figure 4**) was designed and built to test the fully assembled prototype device.

A certified gas standard generator (Kin-Tek FlexStream$^{TM}$ Base Module) was utilized to generate a formaldehyde-air mixture of know concentrations using a permeation tube method. The output formaldehyde level was determined from the output flow rate of gases derived from a compressed air source (~40 psi). To limit the rate of gas flow supplied to the test chamber, the output port of the machine was connected to a pressure relieving regulator, an air-pressure gauge and a 0.01-inch orifice (**Figure 4.2**).

In addition, the formaldehyde fuel cell sensor requires a relative humidity range of 15–95%, while no humidity is allowed in the gas standard generator. Based on this, a custom humidifier was built using a PermSelect® Silicone membrane module (PDMSXA-2500). Air flow was derived from the same compressed air, regulated by a pressure regulator and 0.01-inch orifice, and fed into the humidifier module.

Both the formaldehyde-air mixture and humidified air were fed to one tube for mixing and directed into a test chamber holding the wearable sensor. The test chamber and the humidifier module were placed in a modified temperature-controlled mini-incubator (Benchmark Scientific, H2200-HC, **Figure 4.3**) for temperature control.

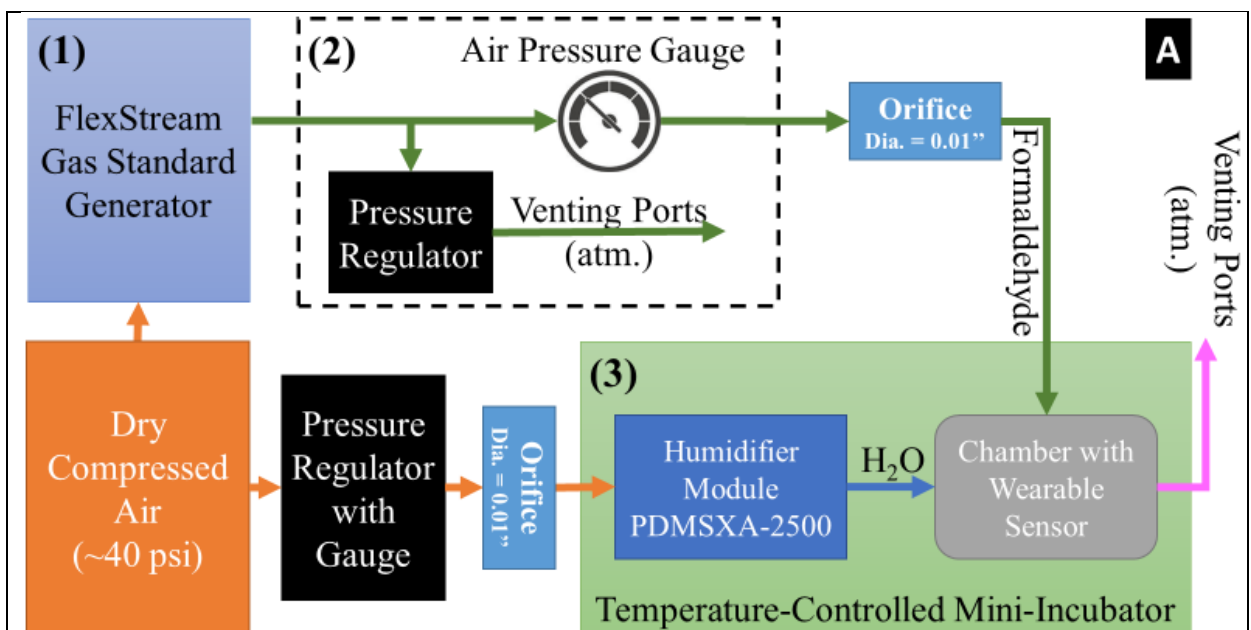

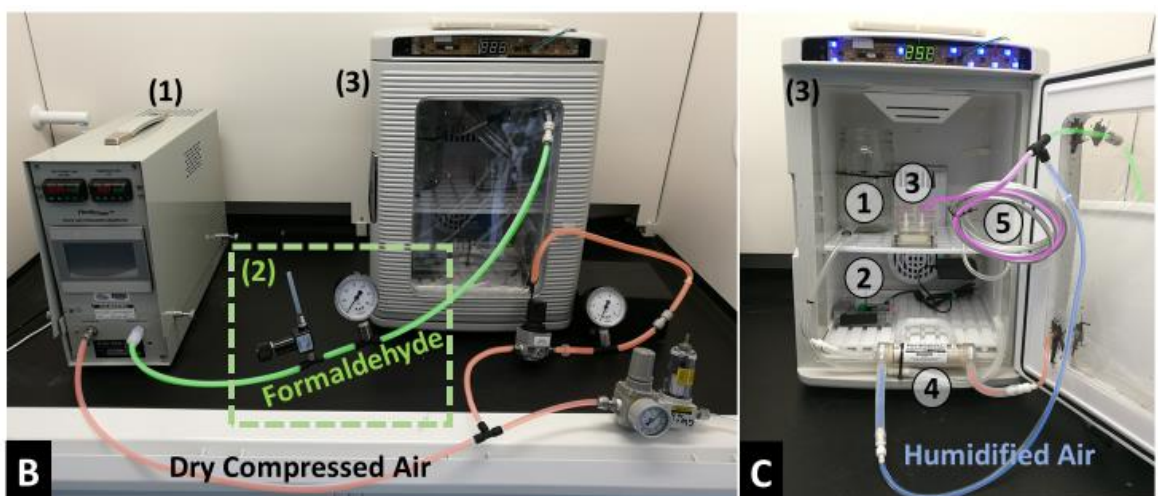

**Figure 4.** Calibration Setup. (A) Block diagram of calibration setup. (B) Overview of the calibration setup in a fume hood. (C) Internal view of the temperature-controlled mini-incubator. ① Pure water to humidify the dry compressed air. ② Raspberry Pi 3 Model B for data acquisition. ③ Test chamber with wearable sensor inside. ④ PermSelect® Silicone membrane module (PDMSXA-2500). ⑤ Long aluminum tubing for gas mixing and equalizing the temperature of the gases generated by the gas standard generator.

## 1.9  Sensor Validation and Reproducibility Tests

Calibration parameters were stored in the on-chip non-volatile flash memory, converting the sensor measurement from ADC readings (arbitrary unit, a.u.) to corresponding ambient aldehyde concentration (parts-per-billion, ppb). Later, two more experiments were conducted to validate the sensor performance; one was performed with standard gas generated at five different concentrations to validate the accuracy of the calibrated sensor measurement, and the other was carried out with standard gas generated at a certain level (206 ppb) for multiple times to examine the reproducibility of the sensor output. **Figure 7** illustrates a sample result from one prototype sensor.

## 2. Results

### 2.1 Electronics System

A standard FR-4 four-layer printed circuit board (PCB) was designed and fabricated with a footprint of 16 x 28 x 1.6 mm (excluding the formaldehyde fuel cell sensor and lithium battery). Once the PCB was received, electronic components were manually populated and tested in the lab.

Firmware was developed using a prototype device built with evaluation modules and breadboards. The firmware image was then loaded into a freshly fabricated sensor. Finally, power consumption was benchmarked using a benchtop power supply (Agilent E3631A) and a digital multimeter (Keithley 2000), typically below 400uA at 4V. Other functions (e.g. *BLE communication, data acquisition, data retrieval, etc.*) were tested with the help of a personal computer.

### 2.2 Fuel Cell Sensor Package

The commercial formaldehyde fuel cell sensor (Dart Sensors Ltd.) comes with a robust package, whose footprint (24 x 27 x 6 mm) is too big to fit into a smartwatch form with other electronic components. However, the core of the fuel cell sensor is two wafers with dimensions of 11 x 11 x 1 mm. These smaller wafers allow re-packaging of the sensor to meet the size requirements.

Custom sensor packages were fabricated with polypropylene (PP) in two pieces (**Figure 5**). Both pieces were milled out of a polypropylene sheet of 1/8-inch thickness using a computer-controlled Benchtop CNC machine (Roland, MDX-40a). A slot of 11.3 x 11.3 x 2.5 mm was milled in the top piece to hold the fuel cell sensors, and a ventilation port of 2 mm in diameter was opened in the center of the slot, permitting ambient air to diffuse to the surface of the sensor. The bottom piece was a rigid polypropylene sheet that matched the outline of the top piece, keeping the fuel cell wafers in place. After assembly, the top and bottom pieces were welded together around the edges using a soldering iron heated to 350ºC.

Two platinum wires were embedded in the enclosure to contact the anode and cathode of the fuel cell. The other ends of the wires were later soldered onto an assembled and tested PCB to form electrical connectivity for data acquisition.

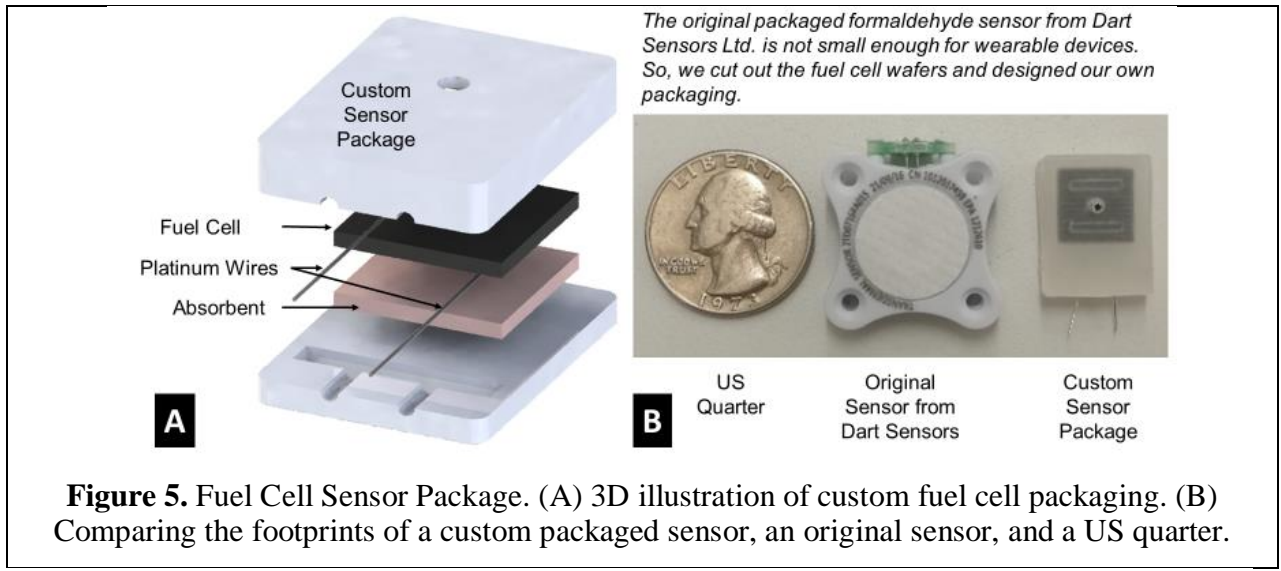

**Figure 5.** Fuel Cell Sensor Package. (A) 3D illustration of custom fuel cell packaging. (B) Comparing the footprints of a custom packaged sensor, an original sensor, and a US quarter.

*2.3 Wearable Device Package and Assembly*

A smart watch enclosure (**Figure 6.A**) with an array of small venting ports (1mm diameter) on the top surface was designed in SolidWorks and prototyped using a high-resolution 3D printer. A porous hydrophobic filter made of polypropylene was attached to the inner surface of the enclosure with adhesives, preventing water from getting into the device and reducing the influence of convection air flows on the fuel cell sensors. The electronic components (PCB, battery, fuel cell sensor) were then assembled into the enclosure. The back cover of the enclosure was then fixed and glued onto the top cover to form good sealing.

A groove was designed on each side of the enclosure, compatible with a wide range of quick release silicone-based bands for smart watches. At the final assembly stage, a pair of smart watch bands were glued onto the wearable device.

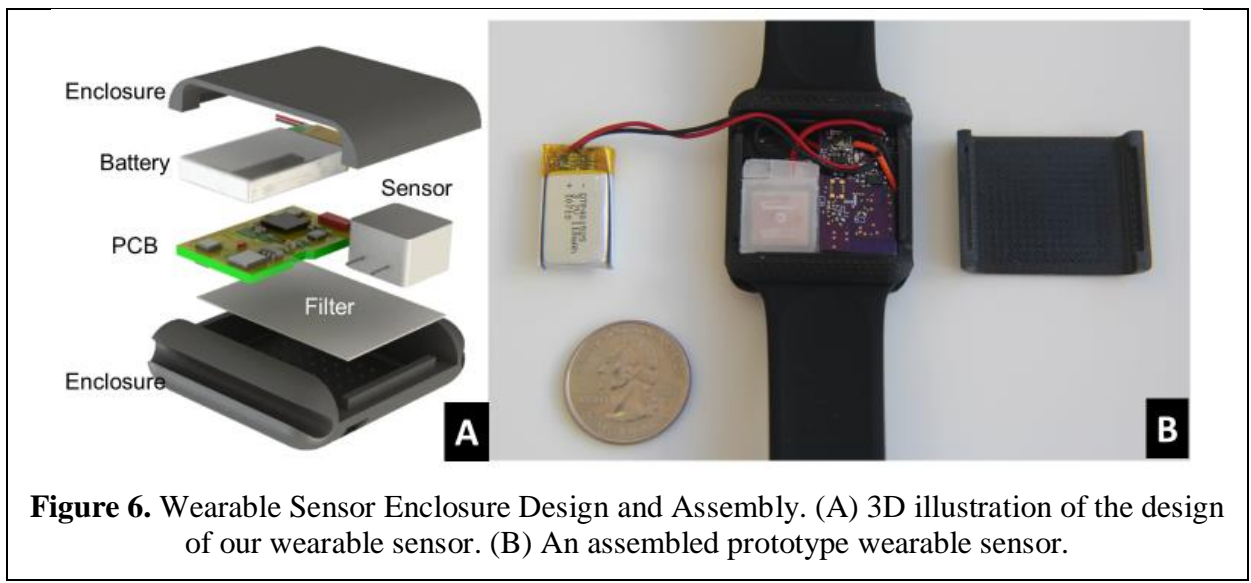

**Figure 6.** Wearable Sensor Enclosure Design and Assembly. (A) 3D illustration of the design of our wearable sensor. (B) An assembled prototype wearable sensor.

Prototype wearable sensors (**Figure 6.B**) were fabricated and tested in a lab environment. The average power consumption measured was approximately 900 µW (3V * 300 µA). The total device weight is about 15g (excluding watch bands).

*2.4   Sensor Calibration and Validation*

After a prototype wearable sensor was assembled and tested, the device was calibrated using the experiment setup described in **Figure 4**. Measurement data (*temperature, relative humidity, and aldehyde level*) was continuously collected by an external single-board Linux-based computer (Raspberry Pi 3 Model B) via BLE wireless communication sampling at a frequency of 1 Hz. The transferred data was then time-stamped and stored in the mini-computer in CSV format. After every experiment, the acquired measurement data was downloaded to a PC and analyzed with MATLAB.

After updating the calibration parameters, sensor validation and reproducibility, experiments were conducted to verify the performance of the calibrated wearable sensor. **Figure 7** shows a sample result from a calibrated prototype device.

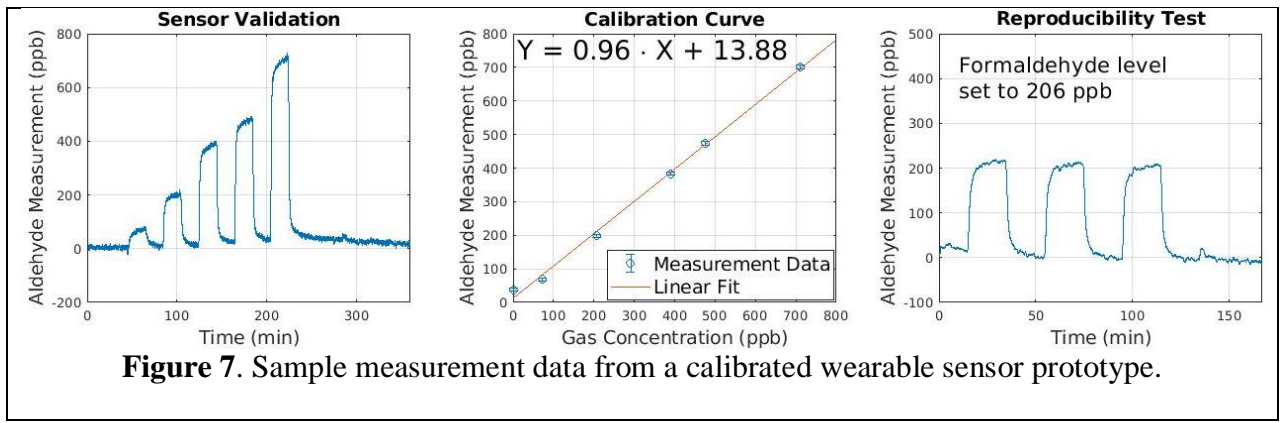

**Figure 7**. Sample measurement data from a calibrated wearable sensor prototype.

*2.5   Smartphone App – User-Interface and Gateway to Cloud*

An Android-based smartphone application (**Figure 8.A**) was developed to demonstrate a proof-of-concept gateway device. This application has four major functions: (1) discover and list nearby BLE-enabled devices, (2) display and plot incoming BLE data packets in real-time when connected to a wearable device, (3) retrieve historical data from the wearable device, and (4) synchronize recorded data with a cloud-based database system for data analytics.

Once a wearable device is connected to the smartphone app, real-time measurement data (*temperature, relative humidity, and formaldehyde level*) will be acquired by the smartphone via BLE characteristics at a frequency of 1 Hz. A button is provided for the user to retrieve historical data stored on the wearable sensor. The downloaded data will be time-stamped, stored in the memory of the smartphone. Three time series graphs, based on an open-source graph plotting library (GraphView), were provided on the graphical user interface (GUI) of the app to display the measurement data.

When internet is available, the stored measurement data will be uploaded to the cloud database system (InfluxDB, details in Section 3.6) using Influxdb-java library [44]. This client library utilizes the InfluxDB HTTP API to communicate with the database server, in which information is organized in JSON format and transmitted to the server over the internet. The app uses the IP address of the database server, username, and password for connection and authentication. During operation, a measurement from one transducer will be inserted as an individual database entry, with its timestamp and other necessary information (details in Section 3.6). The data entries are temporarily stored in an internal buffer in the RAM of the smartphone and flushed to the server as a batch every 10 seconds to reduce communication overhead.

*2.6   Cloud-based Informatics System*

A cloud-based informatics system (**Figure 8.B**) was implemented using Amazon Web Services (AWS). A Ubuntu Server 16.04 image was installed in an Elastic Compute Cloud (EC2) t2.micro instance with an 8GB Elastic Block Storage (EBS) general purpose SSD volume (gp2) for general data storage. This virtual server was managed remotely with secure shell (SSH). Two open-source software packages, *InfluxDB* and *Grafana,* were deployed to store sensor measurement data and provide necessary analytics functions.

*InfluxDB* is an open-source database designed for managing time series data for applications like IoT, DevOps, and data analytics, etc. [45]. Administrator username and password for InfluxDB were configured during installation, and a time-series database for development and demonstration is created afterwards. Every entry in the database represents a measurement from one transducer at a specific time, which consists of *five* essential fields: (1) time-stamp, (2) MAC address of the BLE wearable device, (3) name of the transducer, (4) measurement value, and (5) unit of measurement. The database entries are stored on the EBS volume, allowing the storage space to be scaled as the amount of data increases.

A web-based data visualization and management system was implemented with *Grafana*. It is an open-source software for web-based time series data analytics and monitoring [46,47], which supports various data sources (including InfluxDB) and has a built-in web server with configurable dashboards for data visualization. A dashboard (**Figure 8.C**) was configured to display uploaded measurement data, identified by wearable device ID (BLE MAC address) and time span, in three time plots. This dashboard can refresh at a frequency of 5 seconds to display uploaded data timely.

To track the cloud system status and performance, various metrics of the EC2 instance are monitored using Amazon CloudWatch. Nine metrics involving CPU usage, networking, and storage are monitored: CPUUtilization, CPUCreditUsage, NetworkIn, NetworkOut, StatusCheckFailed_Instance, StatusCheckFailed_System, DiskWriteBytes, and DiskReadBytes. By monitoring StatusCheckFailed, Amazon CloudWatch can automatically restart the EC2 instance to minimize system down-time when it crashes.

EC2 instances can be connected through either its public DNS, public IP, or private IP. When the EC2 instance is relaunched or restarted, these addresses are changed, which can cause problems with re-connecting the informatics system. To address this issue, the EC2 instance is mapped to an Elastic IP address [48], a static public IP address that is automatically mapped to the instance's IP. This ensures that the IP of the instance is always known, including when the instance's IP is changed due to an instance failure. The subdomain prisms.seas.gwu.edu is set to point to the Elastic IP address.

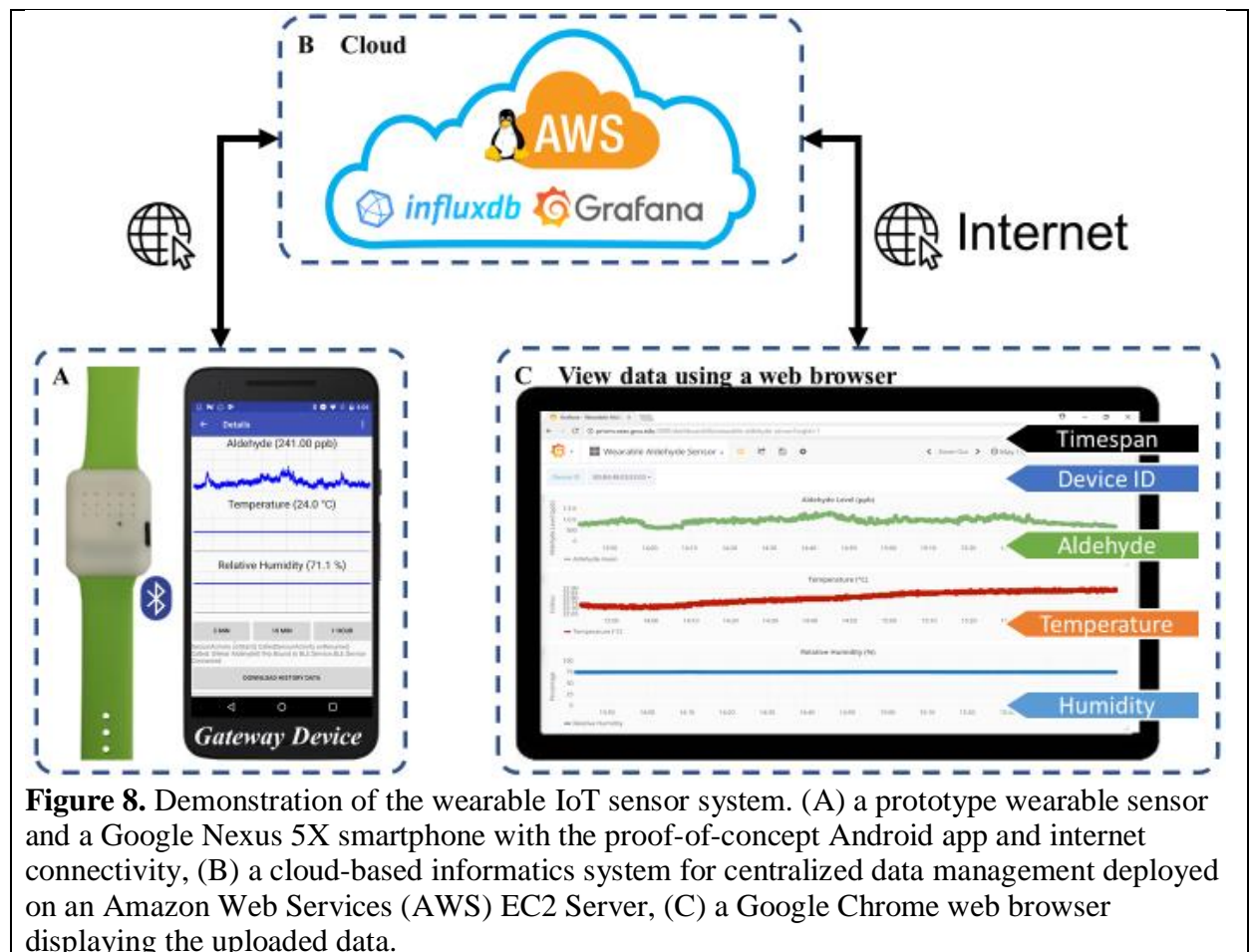

**Figure 8.** Demonstration of the wearable IoT sensor system. (A) a prototype wearable sensor and a Google Nexus 5X smartphone with the proof-of-concept Android app and internet connectivity, (B) a cloud-based informatics system for centralized data management deployed on an Amazon Web Services (AWS) EC2 Server, (C) a Google Chrome web browser displaying the uploaded data.

Our sensors are also compatible with the informatics systems being developed in the NIH PRISMS project [47].

*2.7 Tobacco Smoke Sensing*

One potential application of the wearable aldehyde sensor is environmental tobacco smoke detection and quantification due to the large aldehyde content in tobacco smoke. An experimental setup was built (**Figure 9.A**) to quantify the sensor response to tobacco smoke and study its potential as a tobacco smoke sensor. A 12-Quart plastic storage box with a cover was used as the test chamber. A prototype wearable sensor was placed inside the test chamber with a cigarette and a battery-powered aquarium pump. Tobacco smoke was then generated by burning the cigarette. A constant air flow was provided by the pump to pass through the cigarette to facilitate the burning process. The test chamber was placed in a chemical fume hood and a 1/4-inch diameter hole was drilled on the cover of the chamber for ventilation.

**Figure 9.B** shows typical measurement data from one such tobacco smoke sensing test. The cigarette was consumed in 1 minute after the experiment began, and the sensor output reached its peak value in about 5 minutes, followed by an exponential decay as tobacco smoke diffused out of the chamber.

The sensor reading reached around 600 ppm at its peak, which is equivalent to 200 ppb in a 140 square-feet room (assuming 2.5 meters tall). In real-world settings, ventilation and air-conditioning may significantly lower this concentration. However, one previous study showed that formaldehyde concentration can reach 250ppb in a poorly ventilated room after 10-15 cigarette smoke [9], which is in the detectable range of our sensor.

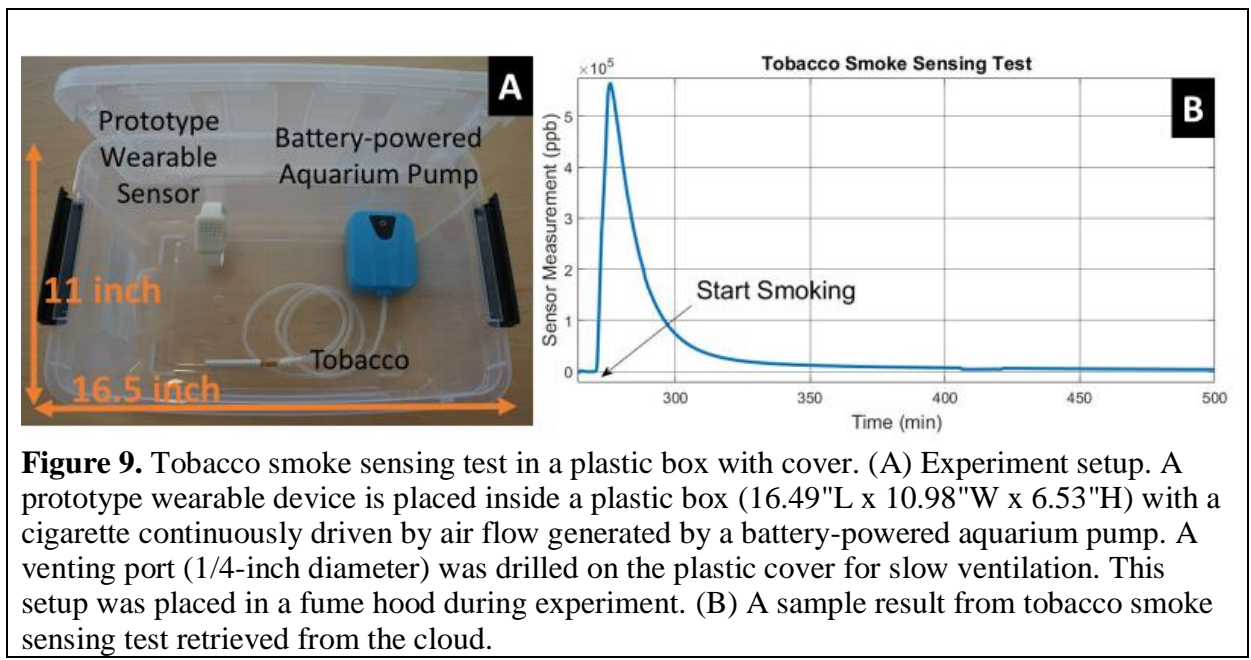

**Figure 9.** Tobacco smoke sensing test in a plastic box with cover. (A) Experiment setup. A prototype wearable device is placed inside a plastic box (16.49"L x 10.98"W x 6.53"H) with a cigarette continuously driven by air flow generated by a battery-powered aquarium pump. A venting port (1/4-inch diameter) was drilled on the plastic cover for slow ventilation. This setup was placed in a fume hood during experiment. (B) A sample result from tobacco smoke sensing test retrieved from the cloud.

## 3. Discussion

Specificity has been one of the major challenges for low-power miniature gas sensors [49]. The formaldehyde fuel cell sensor has measurable responses to various aldehydes and other interference substances, such as alcohol and $SO_2$. Its specificity is mainly determined by the catalysts coated on the surfaces of the fuel cell. Other types of miniature gas sensors, such as metal oxide sensors, also have specificity issues [50]. Although various techniques have been developed to increase the specificity and selectivity of gas sensors [49–52], to our knowledge, it is still one of the biggest challenges for miniature gas sensors.

One possible method to improve specificity is to incorporate multiple sensing modalities working under different sensing principles to increase the dimensions of the measurement data and apply signal processing techniques (e.g. principle component analysis) to distinguish and quantify various gas contents [53,54].

Although our custom packaged fuel cell sensor was calibrated with a standard gas generator and the calibration results were consistent across various experiments, the wearable sensors exhibited baseline fluctuations during real-world on-wrist tests. Four possible causes may contribute to these fluctuations: (1) the fluctuation of a temperature gradient across the fuel cell sensor thickness [55], (2) movement-induced air convection, (3) motion artifacts, and (4) electromagnetic interference (EMI). The first one appeared to have the most significant effect in our field tests. This thermal gradient variation and potential EMI can be solved by encapsulating the fuel cell sensor in a grounded metallic enclosure. The thermal gradient effect can also be compensated in software by measuring it with two temperature sensors. Sources (2) and (3) can be reduced by smaller pore size membrane filters and more secure mechanical fit designs, respectively.

The fuel cell sensors have transient responses to fast changes in ambient relative humidity. Our experiment results showed that it generally took 1-5 min for the sensor to recover. According to the fuel cell manufacturer, this is likely caused by thermal effects, where excessive water molecules dissolved into the acid-containing proton exchange membrane and generated heat. These transient responses can be removed in post-processing, based on relative humidity data. Further studies are required to validate if such transient responses can be corrected with the relative humidity measurements.

We also realized that some of the chemical residue on the circuit boards introduced during fabrication can evaporate and interact with the fuel cell sensors, generating a baseline shift. Also, the releasing rate depends on temperature, which makes the baseline shift change significantly with temperature. We reduced this baseline shift by soaking the PCB in ethanol (99%) for 1 hour, thoroughly brushing it under running pure water, and drying it using compressed air. In future designs, it would be preferable to have the sensor air pathway directly connected to the ambient environment to minimize the chances of false readings.

Despite these challenges, fuel cell sensors have many advantages that make them one of the most promising sensing modalities for wearable gas monitors. Take the formaldehyde sensor from Dart Sensors as an example. The sensing element has a small footprint (11 x 11 x 2 mm) and does not require a power supply for operation. The lower limit of detection (LLOD) can reach 15-30 ppb in a well-controlled (lab) environment. The performance of each individual sensor is consistent over an extended period (months). Further, the platinum-based fuel cell sensor only responds to substances that can be reduced or oxidized at a platinum electrode, which intrinsically provides certain specificity [56]. In our field tests, the major source of interference observed was due to various types of alcohols, which can potentially be distinguished by adding other sensing modalities and using blind signal separation techniques [57].

When combined with other wearable environmental sensors [58–61] and physiological sensors such as a wheezing detection patch or a spirometer, this wearable IoT sensor platform can provide objective exposure data correlated with asthma symptoms and physiological responses. It can also be used in a citizen-science environmental monitoring system to map air pollution at much higher spatial and temporal resolutions than currently available. Other applications include new house and vehicle inspections and furniture selections which may benefit from on-site formaldehyde sensing.

## 4. Conclusion

In this paper, we describe a personalized wearable IoT sensor platform for monitoring ambient aldehyde level. A self-contained wearable aldehyde sensing device with the size comparable to a smart watch was designed and built using commercial aldehyde fuel cell sensors and state-of-the-art low-power electronics. The average power consumption measured was 900μW, enabling the device to last over a week with a fully charged 100mAh lithium battery. The prototype devices were calibrated in the laboratory with a permeation tube based standard gas generator. The lower limit of detection (LLOD) was 30ppb and upper detection limit was larger than 10ppm. A dedicated Android App was developed on a Google Nexus 5X smartphone, serving as a gateway to wirelessly retrieve sensor data and upload it to a cloud-based informatics system implemented on AWS for storage, visualization and analysis. Further engineering improvements are necessary to minimize the fluctuations of the fuel cell sensor due to temperature gradient changes in on-body deployments.

We envision this wearable environmental IoT sensor platform will be used by physicians and researchers to continuously monitor the aldehyde exposure level of asthma patients over time. The aldehyde exposure levels can be correlated with asthma attack events and various symptoms to discover any possible relationship. In the future, this system can potentially be applied for personalized asthma management if a causal relation is identified between aldehyde exposure and asthma exacerbations for a given individual.

**Acknowledgements**
This research was supported by the National Institute of Biomedical Imaging and Bioengineering of the National Institutes of Health under award number U01EB021986. The authors thank Dr. Walter King, Dr. Katherine Sward, Dr. Neal Patwari, and Philip Lundrigan for helpful discussions.